%% file: main.tex
\documentclass[amsmath, amsthm, amssymb, floatfix, reprint, twocolumn, notitlepage, nofootinbib, 10pt]{revtex4-1}
\usepackage{setspace}
\usepackage{amsmath}
\usepackage{breqn}
\usepackage{graphicx}
\usepackage[nearskip,margin = 0pt,caption=false]{subfig}

\usepackage{verbatim}
\usepackage{amsfonts}
\usepackage{amssymb}
\usepackage{epstopdf}
\usepackage{lipsum}
\usepackage{siunitx}
\usepackage{xcolor}
\usepackage{array}
\usepackage{braket}
\usepackage[normalem]{ulem}
\usepackage[resetlabels,labeled]{multibib}
\DeclareGraphicsExtensions{.pdf,.eps,.png,.jpg,.mps}
\usepackage{etoolbox}
\usepackage{subfiles} 

\begin{document}
\title{
Gradient-Based Optimization of Optical Vortex Beam Emitters}
\author{Alexander D. White$^{1, *}$, Logan Su$^1$, Daniel I. Shahar$^2$, Ki Youl Yang$^1$, Geun Ho Ahn$^1$, \\ Jinhie Skarda$^1$, Siddharth Ramachandran$^2$, Jelena Vu\v{c}kovi\'{c}$^{1, *}$\\
\vspace{+0.05 in}
$^1$E. L. Ginzton Laboratory, Stanford University, Stanford, CA 94305, USA.\\
$^2$Department of Electrical and Computer Engineering, Boston University, Boston, Massachusetts 02215, USA.\\
{\small $*$ adwhite@stanford.edu, jela@stanford.edu}}

\begin{abstract}
    \noindent
    Vortex beams are stable solutions of Maxwell's equations that carry phase singularities and orbital angular momentum, unique properties that give rise to many applications in the basic sciences, optical communications, and quantum technologies.  
    Scalable integration and fabrication of vortex beam emitters will allow these applications to flourish and enable new applications not possible with traditional optics.
    Here we present a general framework to generate integrated vortex beam emitters using photonic inverse design. We experimentally demonstrate generation of vortex beams with angular momentum spanning -3$\hbar$ to 3$\hbar$. We show the generality of this design procedure by designing a vortex beam multiplexer capable of exciting a custom vortex beam fiber. Finally, we produce foundry-fabricated beam emitters with wide-bandwidths and high-efficiencies that take advantage of a multi-layer heterogeneous integration.
\end{abstract}

\maketitle

Optical vortex beams and the orbital angular momentum (OAM) they carry are useful in a vast array of applications including classical and quantum communication, super-resolution microscopy, optical trapping and manipulation, and metrology \cite{ni2021multidimensional,shen2019optical, barreiro2008beating,wang2012terabit, ren2016chip}. While these beams are traditionally generated by phase plates or spatial light modulators, efforts have been made to integrate optical vortex beams emitters onto nanophotonic platforms. Such integration allows for the construction of extremely compact vortex beam generators in platforms that can provide additional functionality and optical processing.

Initial demonstrations of on-chip vortex beam generators consisted of ring resonators with gratings whose periodicity mismatch with the optical mode enforced the radial phase required for an OAM beam \cite{cai2012integrated}. While this design is capable of producing high quality OAM beams in a compact area, the resonant nature of the devices limit their bandwidth and prevent multiplexing. Since then, many strategies for on-chip vortex beam generation have been demonstrated including OAM lasers \cite{miao2016orbital, liu2017wavelength, kitamura2019generation, bahari2021photonic}, metasurfaces \cite{yu2011light, chen2018orbital, sedeh2020time}, and analytical \cite{su2012demonstration, zhou2019ultra, zhao2020design} and computational \cite{ xie2018ultra, song2020utilizing, white2021inverse} grating design.

Here we provide a general framework for the design of integrated vortex beam emitters using adjoint optimized photonics inverse design. We first demonstrate OAM beam generation of $\ell = -3$ to $3$ in single layer silicon photonics. We then show a 3 mode OAM multiplexer designed to launch modes into a custom $\ell = -1, 0, 1$ fiber. Finally, we utilize the well controlled fabrication and heterogeneous integration of a commercial foundry (A*STAR, AMF) to demonstrate high fidelity, wide bandwidth, and efficient vortex beam emitters. While we focus on vortex beam generation here, the same inverse design protocol can be used to optimize beam emitters and multiplexers for arbitrary free space modes.

Vortex beams consist of a Laguerre-Gaussian field profile with a phase that is linearly proportional to the angle and integrates to 2$\pi$ times an integer $\ell$, the orbital angular momentum:
\begin{equation}
    E_{p,\ell}(r, \phi) = \frac{1}{w_0} L_p^{|\ell|}\left(\frac{2r^2}{w_0^2}\right) \sqrt{\frac{2p!}{\pi(p+|\ell|)!}} \left(\frac{r\sqrt{2}}{w_0}\right)^{|\ell|} e^{\frac{-r^2}{w_0^2}} e^{i\ell\phi},
\end{equation}
where $p$ is the radial index and $w_0$ is the beam waist. For first order radial modes, the simplest OAM beams, this profile simplifies to
\begin{equation}
    E_{\ell}(r, \phi) = \frac{1}{w_0} \sqrt{\frac{1}{\pi|\ell|!}} \left(\frac{r\sqrt{2}}{w_0}\right)^{|\ell|} e^{\frac{-r^2}{w_0^2}} e^{i\ell\phi},
\end{equation}
examples of which are shown in Figure 1a. Vortex beams can have nearly arbitrary polarization, including topologically interesting spatially varying polarization coupled with the beam shape \cite{bauer2015observation}, but here we use spatially homogeneous linear polarization for simplicity.
    
\begin{figure*}[b!]
\centering
\includegraphics[width=0.95\linewidth]{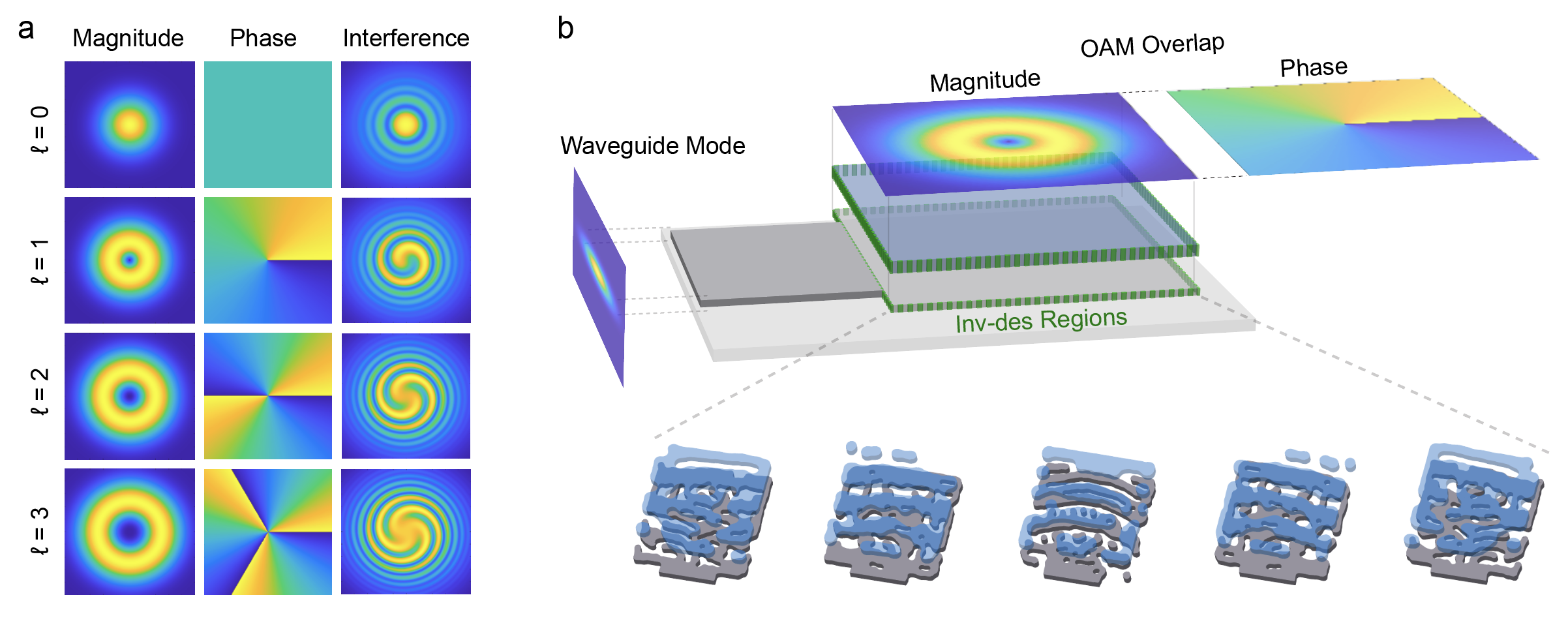}
\caption{\label{fig:Fig1}{\bf{Inverse Design Approach}} \textbf{(a)} Cross-section of vortex beams with OAM $\ell$ from 0 to 3. Plots show magnitude, phase, and the interference pattern generated by interfering with a Gaussian beam. \textbf{(b)} Schematic of the inverse design approach to generating optical vortex beam emitters. We optimize an area (Inv-des Regions) that takes a waveguide mode as input and generates a field pattern that has a maximum overlap with the desired OAM beam. Grey layers show silicon and blue show silicon nitride.}
\end{figure*}

\begin{figure*}[t!]
\centering
\includegraphics[width=0.95\linewidth]{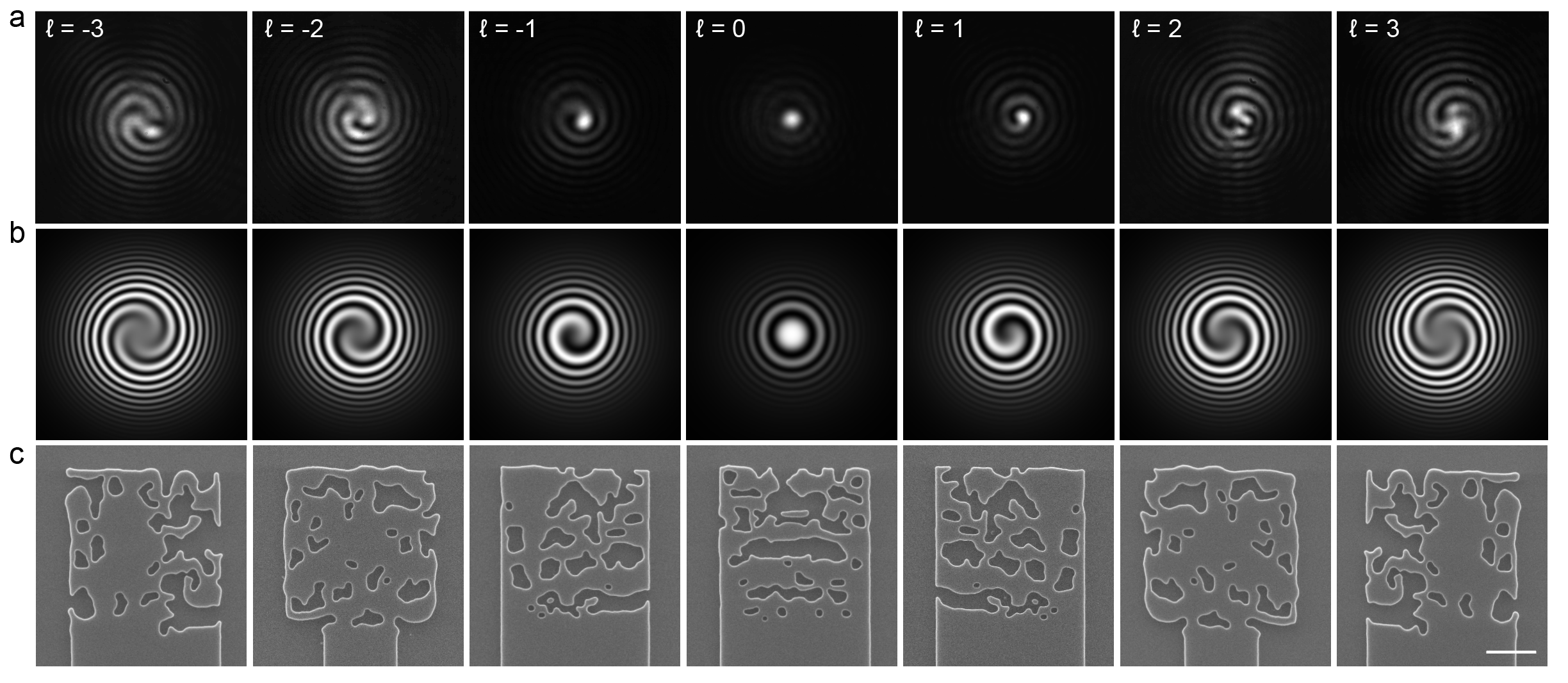}
\caption{\label{fig:Fig2}{\bf{Single Layer Vortex Beam Emitters}} \textbf{(a)} Measured interference patterns generated from single layer silicon OAM gratings driven with a 1515~nm laser through the waveguide and a co-linear Gaussian beam. \textbf{(b)} Simulated interference patterns corresponding to the patterns in (a). \textbf{(c)} SEM images of the vortex beam emitter devices used in (a). Scale bar is 1$\mu$m.}
\end{figure*}

\begin{figure*}[t!]
\centering
\includegraphics[width=0.95\linewidth]{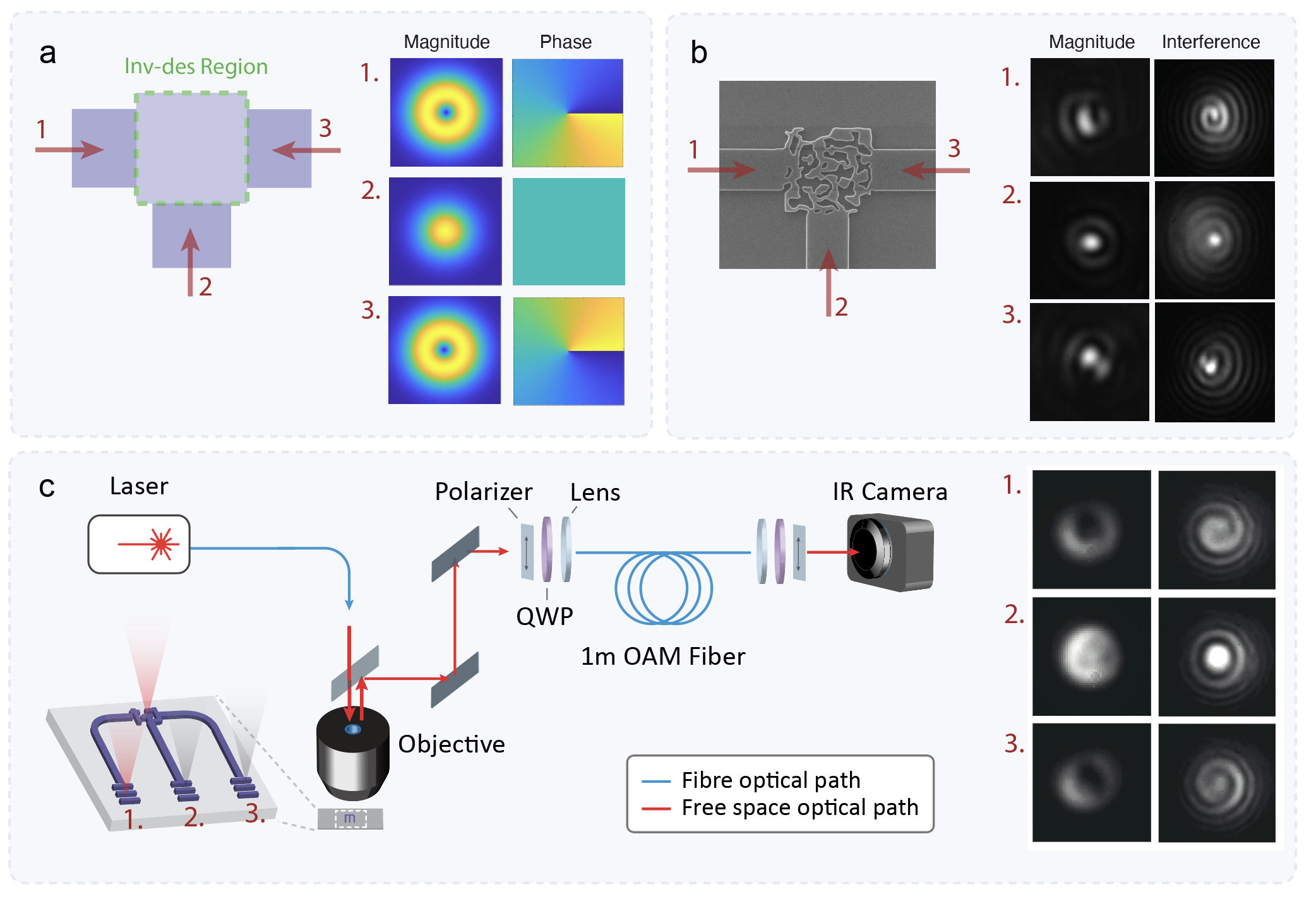}
\caption{\label{fig:Fig3}{\bf{Vortex Beam Multiplexer}} \textbf{(a)} Schematic of the vortex beam multiplexer design. The inverse design region has 3 waveguide inputs and is optimized to launch $\ell = -1, 0, 1$ from ports 1, 2, and 3 respectively. In order to efficiently multiplex these beams, all are designed to emit linearly polarized in the y (vertical) axis. \textbf{(b)} SEM and measured intensity and interference patterns from the fabricated device, measured at 1510~nm. \textbf{(c)} Schematic of fiber coupling measurement and measured patterns. The y-polarized beams are sent through a quarter wave plate (QWP) before and after the fiber in order to couple to the circularly polarized eigenmodes of the fiber.}
\end{figure*}

To design arbitrary beam emitters, we optimize a structure that takes a waveguide mode as input and launches a beam out of the plane of the chip whose electric field overlaps maximally with the desired beam shape in free space (Figure 1b). To efficiently optimize the structure, we use an adjoint optimization approach which allows us to calculate the optimization gradients at every point with only two simulations \cite{goos, su2020nanophotonic}. Here we optimize the emitters to form vortex beams, but this design approach is general and can be applied to any desired spatial field pattern. 

We first optimize compact devices that emit vortex beams for $\ell = -3$ to $3$. These devices are $3 \times 3$~$\mu$m and fabricated using electron beam lithography on an air-clad 220~nm silicon-on-insulator platform. They operate in the telecom band, and are designed with an 80~nm minimum feature size. To measure the emission properties, we excite an on-chip waveguide coupled to a device and collect the emitted light through a high NA objective. We then interfere the beam from the device with a Gaussian beam generated from the same laser source and image the resulting intensity pattern. Figure 2 shows the theoretical interference patterns from ideal vortex beams as well as the measured interference patterns (with simulations shown in Supplemental Figure 1). The spiral pattern is formed by the combination of a curved phase front radially and the angular OAM phase; the direction of spiraling shows the sign of the OAM and the number of teeth show the OAM order. 

\begin{figure*}[t!]
\centering
\includegraphics[width=0.95\linewidth]{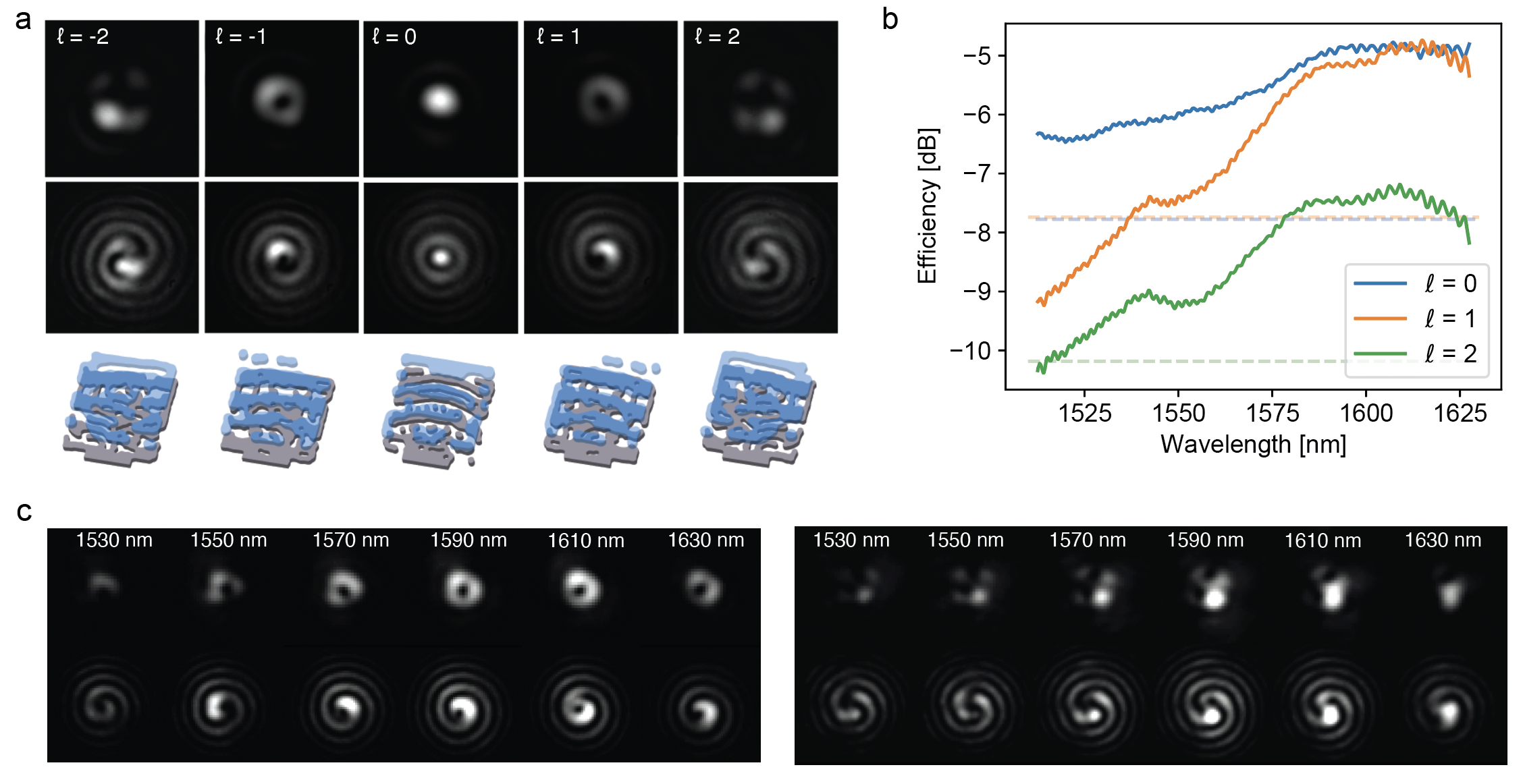}
\caption{\label{fig:Fig4}{\bf{Foundry-fabricated 2-layer Beam Emitters}} \textbf{(a)} Intensity (top row) and interference (center row) patterns from foundry fabricated 2-layer beam emitters measured at 1570~nm. Bottom row shows 3D representations of the grating devices with silicon in grey and silicon nitride in blue. \textbf{(b)} Efficiency of gratings measured using a 50um core multi-mode fiber; $\ell = -1, -2$ efficiencies are identical to $\ell = 1, 2$, as the devices are mirror images of each other. Dashed lines show 3dB from each maxima. \textbf{(c)} Wavelength sweep of $\ell = 1$ and $\ell = 2$ intensity and interference patterns from 1530 to 1630~nm.}
\end{figure*}

To demonstrate the generality of this design approach, we optimize a device to launch multiplexed $\ell = -1,0,1$ modes into a custom OAM supporting optical fiber \cite{bozinovic2013terabit}. The device is designed to launch an $\ell = -1$ mode when excited from the left waveguide, $\ell = 1$ when excited from the right, and $\ell = 0$ when excited from the bottom waveguide (Figure 3a). Additionally, all of these modes are designed to be y-polarized so that a polarizer can be used to filter out undesired mode coupling in the fiber. This leads to the additional design complication that the $\ell = 0$ mode must be launched with the orthogonal polarization to its input waveguide mode. Even with these additional constraints, the optimization procedure yielded a compact $3 \times 3$~$\mu$m device that was able to generate all 3 modes (Figure 3b, Supplemental Figure 2). To launch these modes into a custom OAM fiber, we pass the output fields through a polarizer and quarter wave plate to match the circularly polarized eigenmodes of the fiber, and use a lens to focus the collimated beams on the fiber core. To image the fiber modes with a camera, we employ another lens, quarter wave plate, and polarization beamsplitter (Figure 3c). This leads to beams which are even more pure than those directly from the multiplexer, as the fiber filters out any higher order modes from design nonidealities and scattering.

Finally, we utilize the well controlled fabrication and heterogeneous integration from the A*STAR AMF foundry to generate high fidelity, wide bandwidth and efficient vortex beam emitters. To do this, we implement an FDFD adjoint optimization package capable of the simultaneous optimization of multiple design regions \cite{goos}. In combination with multilayer photonics processing \cite{sacher2015multilayer}, this allows us to design fully 3D structures parameterized by slices in the $z$ dimension. We design $\ell = -2$ to $2$ devices consisting of two co-optimized layers; a 220~nm silicon layer with a 140~nm minimum feature size, and a 400~nm silicon nitride layer with a 300~nm minimum feature size. The layers are separated by 250~nm and are clad with silicon oxide. 

The measured intensity and interference patterns of these devices is shown in Figure 4a (simulated results shown in Supplemental Figure 3). The additional degrees of freedom given by the 2 layer design, as well as the better controlled fabrication process of a foundry, allows us to see significant improvements in the beam shapes. Additionally, these emitters are able to maintain OAM emission across at least a 100~nm optical bandwidth (Figure 4b). To measure the emitter efficiency, we couple the to a multimode fiber that supports OAM beams and sweep the input wavelength and find that the 3dB bandwidths exceed 100~nm (Figure 4c).

We demonstrated a general optimization technique for beam emitters and applied it to generate optical vortex beams. We showed compact generation of $\ell = -3$ to $3$ vortex beams, a vortex beam multiplexer, and high-performance foundry-fabricated heterogeneously-integrated multi-layer devices. The robustness of this approach in generating multiple orders and combinations of vortex beams and its ability to translate to foundry processing opens the door for applications in classical and quantum communications, information processing, and imaging.

\noindent\textbf{Materials and Methods}
\\
\noindent\textbf{Optimizations}
Using adjoint optimization as described in \cite{su2020nanophotonic}, we maximize the free space overlap of OAM beams given a fundamental TE waveguide mode input. This maximization is defined as 
\begin{equation*}
    \max_\varepsilon |\text{c}^\dagger \text{x}(\varepsilon)|^2
\end{equation*}
where $\text{x}(\varepsilon)$ is the vectorized electric field due to the designed permitted distribution and $\text{c}$ is the vectorized electric field of the desired OAM beam in free space. The devices in figures 2 and 3 were optimized for 1550~nm, and the devices in figure 4 were optimized for both 1530~nm and 1570~nm. The software to perform these optimization is publicly available at \cite{goos}.

\bibliography{main.bbl}

\vspace{150pt}

\noindent\textbf{Acknowledgments}
The authors thank Kasper Van Gasse and Rahul Trivedi for productive discussions, and thank Mohamad Idjadi and Firooz Aflatouni for assistance in foundry integration. This research is supported by the Air Force Office of Scientific Research (FA9550-17-1-0002), DARPA under the PIPES program, the Vannevar Bush Faculty Fellowship (N00014-19-1-2632), and ONR-MURI (N00014-20-1-2450).
\noindent

\clearpage 



\renewcommand{\thefigure}{S\arabic{figure}}
\setcounter{figure}{0}

\pagebreak

\subfile{supplement.tex}

\end{document}

%% file: supplement.tex
\title{Supplemental Information}

\begin{figure*}[h!]
\centering
\includegraphics[width=0.9\linewidth]{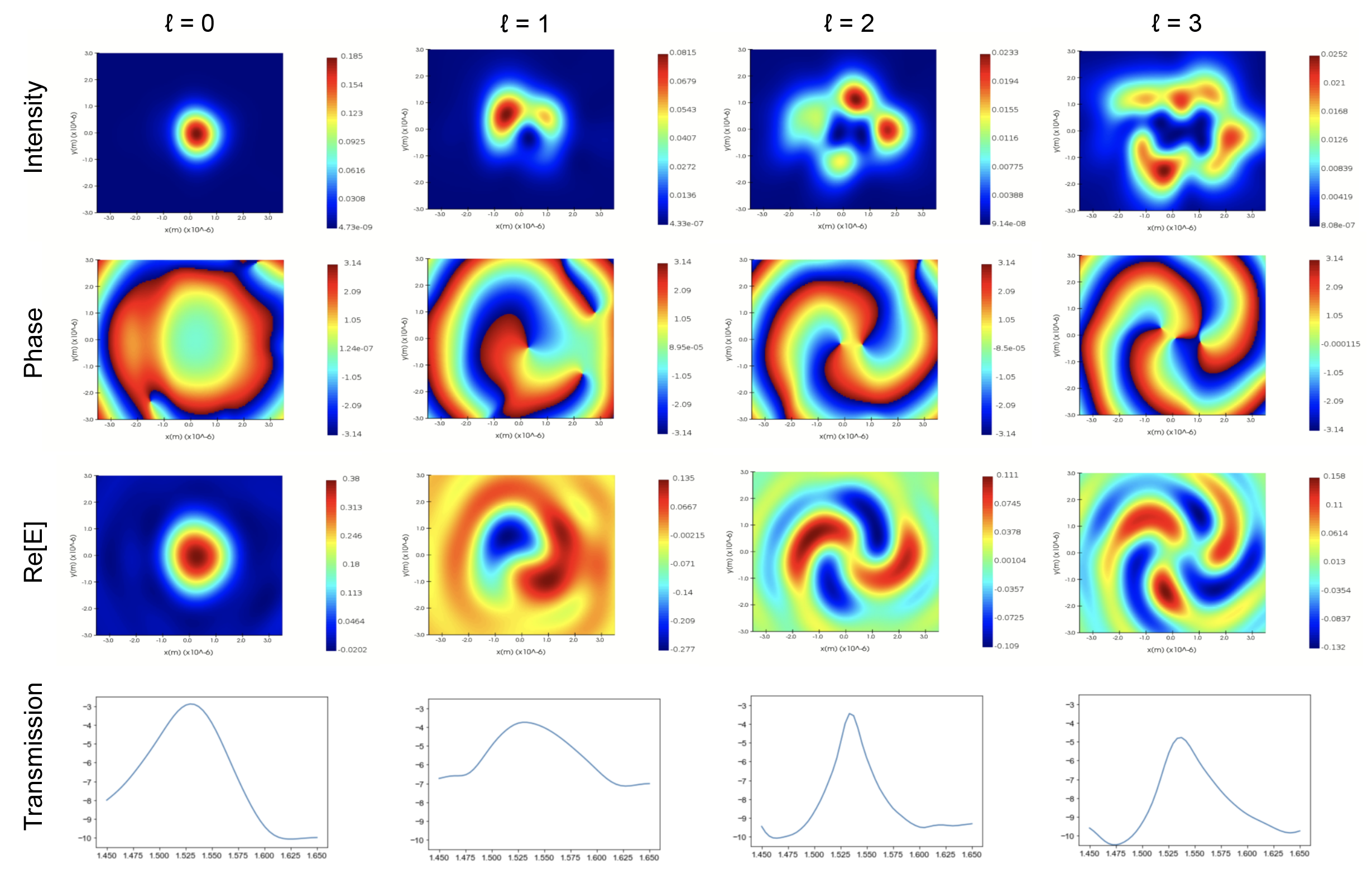}
\caption{\label{fig:Fig2}{\bf{Simulated Single Layer Vortex Beam Emitters}} Simulations of intensity, phase, and real part of the laterally-polarized electric field for single layer couplers at 1545nm. Bottom row shows corresponding transmission spectra. }
\end{figure*}

\begin{figure*}[h!]
\centering
\includegraphics[width=0.7\linewidth]{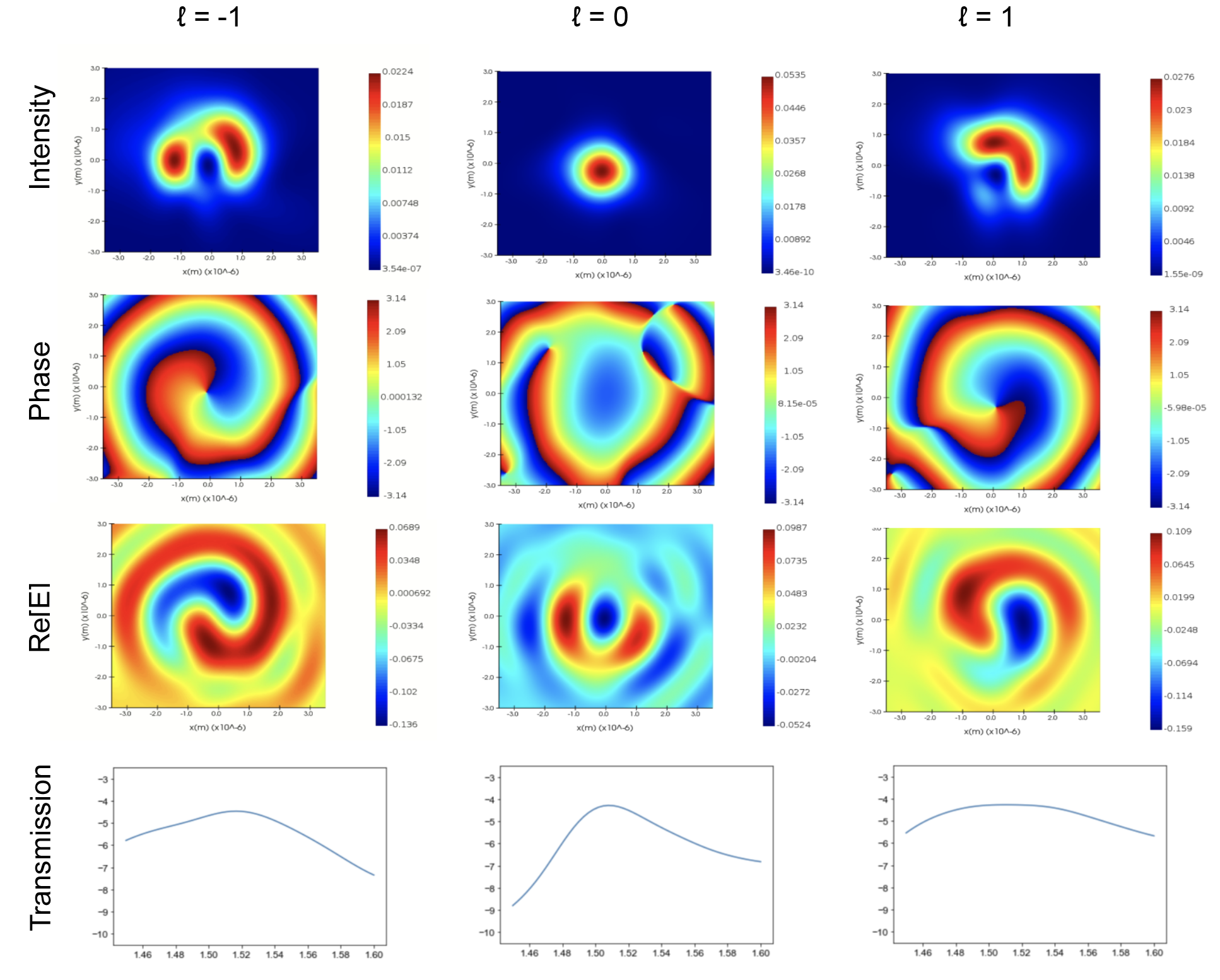}
\caption{\label{fig:Fig2}{\bf{Simulated Vortex Beam Multiplexer}} Simulations of intensity, phase, and real part of the Y-polarized electric field for OAM multiplexer at 1522nm. Bottom row shows corresponding transmission spectra. }
\end{figure*}

\begin{figure*}[h!]
\centering
\includegraphics[width=0.7\linewidth]{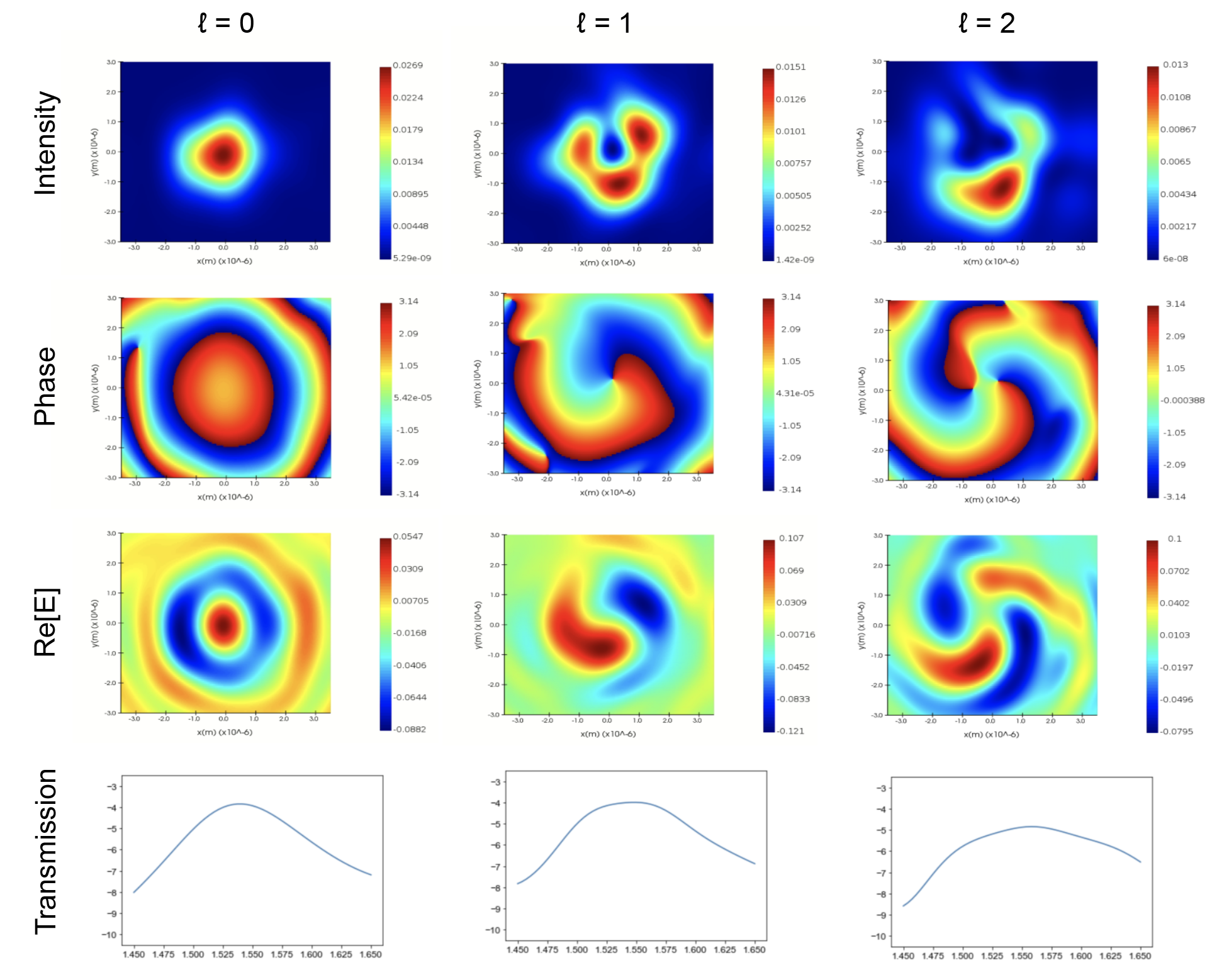}
\caption{\label{fig:Fig2}{\bf{Simulated 2-Layer Vortex Beam Emitters}} Simulations of intensity, phase, and real part of the laterally-polarized electric field for two-layer couplers at 1550nm. Bottom row shows corresponding transmission spectra. }
\end{figure*}

%% file: main.bbl
\begin{thebibliography}{24}%
\makeatletter
\providecommand \@ifxundefined [1]{%
 \@ifx{#1\undefined}
}%
\providecommand \@ifnum [1]{%
 \ifnum #1\expandafter \@firstoftwo
 \else \expandafter \@secondoftwo
 \fi
}%
\providecommand \@ifx [1]{%
 \ifx #1\expandafter \@firstoftwo
 \else \expandafter \@secondoftwo
 \fi
}%
\providecommand \natexlab [1]{#1}%
\providecommand \enquote  [1]{``#1''}%
\providecommand \bibnamefont  [1]{#1}%
\providecommand \bibfnamefont [1]{#1}%
\providecommand \citenamefont [1]{#1}%
\providecommand \href@noop [0]{\@secondoftwo}%
\providecommand \href [0]{\begingroup \@sanitize@url \@href}%
\providecommand \@href[1]{\@@startlink{#1}\@@href}%
\providecommand \@@href[1]{\endgroup#1\@@endlink}%
\providecommand \@sanitize@url [0]{\catcode `\\12\catcode `\$12\catcode
  `\&12\catcode `\#12\catcode `\^12\catcode `\_12\catcode `\%12\relax}%
\providecommand \@@startlink[1]{}%
\providecommand \@@endlink[0]{}%
\providecommand \url  [0]{\begingroup\@sanitize@url \@url }%
\providecommand \@url [1]{\endgroup\@href {#1}{\urlprefix }}%
\providecommand \urlprefix  [0]{URL }%
\providecommand \Eprint [0]{\href }%
\providecommand \doibase [0]{http://dx.doi.org/}%
\providecommand \selectlanguage [0]{\@gobble}%
\providecommand \bibinfo  [0]{\@secondoftwo}%
\providecommand \bibfield  [0]{\@secondoftwo}%
\providecommand \translation [1]{[#1]}%
\providecommand \BibitemOpen [0]{}%
\providecommand \bibitemStop [0]{}%
\providecommand \bibitemNoStop [0]{.\EOS\space}%
\providecommand \EOS [0]{\spacefactor3000\relax}%
\providecommand \BibitemShut  [1]{\csname bibitem#1\endcsname}%
\let\auto@bib@innerbib\@empty
\bibitem [{\citenamefont {Ni}\ \emph {et~al.}(2021)\citenamefont {Ni},
  \citenamefont {Huang}, \citenamefont {Zhou}, \citenamefont {Gu},
  \citenamefont {Song}, \citenamefont {Kivshar},\ and\ \citenamefont
  {Qiu}}]{ni2021multidimensional}%
  \BibitemOpen
  \bibfield  {author} {\bibinfo {author} {\bibfnamefont {J.}~\bibnamefont
  {Ni}}, \bibinfo {author} {\bibfnamefont {C.}~\bibnamefont {Huang}}, \bibinfo
  {author} {\bibfnamefont {L.-M.}\ \bibnamefont {Zhou}}, \bibinfo {author}
  {\bibfnamefont {M.}~\bibnamefont {Gu}}, \bibinfo {author} {\bibfnamefont
  {Q.}~\bibnamefont {Song}}, \bibinfo {author} {\bibfnamefont {Y.}~\bibnamefont
  {Kivshar}}, \ and\ \bibinfo {author} {\bibfnamefont {C.-W.}\ \bibnamefont
  {Qiu}},\ }\href@noop {} {\bibfield  {journal} {\bibinfo  {journal} {Science}\
  }\textbf {\bibinfo {volume} {374}},\ \bibinfo {pages} {eabj0039} (\bibinfo
  {year} {2021})}\BibitemShut {NoStop}%
\bibitem [{\citenamefont {Shen}\ \emph {et~al.}(2019)\citenamefont {Shen},
  \citenamefont {Wang}, \citenamefont {Xie}, \citenamefont {Min}, \citenamefont
  {Fu}, \citenamefont {Liu}, \citenamefont {Gong},\ and\ \citenamefont
  {Yuan}}]{shen2019optical}%
  \BibitemOpen
  \bibfield  {author} {\bibinfo {author} {\bibfnamefont {Y.}~\bibnamefont
  {Shen}}, \bibinfo {author} {\bibfnamefont {X.}~\bibnamefont {Wang}}, \bibinfo
  {author} {\bibfnamefont {Z.}~\bibnamefont {Xie}}, \bibinfo {author}
  {\bibfnamefont {C.}~\bibnamefont {Min}}, \bibinfo {author} {\bibfnamefont
  {X.}~\bibnamefont {Fu}}, \bibinfo {author} {\bibfnamefont {Q.}~\bibnamefont
  {Liu}}, \bibinfo {author} {\bibfnamefont {M.}~\bibnamefont {Gong}}, \ and\
  \bibinfo {author} {\bibfnamefont {X.}~\bibnamefont {Yuan}},\ }\href@noop {}
  {\bibfield  {journal} {\bibinfo  {journal} {Light: Science \& Applications}\
  }\textbf {\bibinfo {volume} {8}},\ \bibinfo {pages} {1} (\bibinfo {year}
  {2019})}\BibitemShut {NoStop}%
\bibitem [{\citenamefont {Barreiro}\ \emph {et~al.}(2008)\citenamefont
  {Barreiro}, \citenamefont {Wei},\ and\ \citenamefont
  {Kwiat}}]{barreiro2008beating}%
  \BibitemOpen
  \bibfield  {author} {\bibinfo {author} {\bibfnamefont {J.~T.}\ \bibnamefont
  {Barreiro}}, \bibinfo {author} {\bibfnamefont {T.-C.}\ \bibnamefont {Wei}}, \
  and\ \bibinfo {author} {\bibfnamefont {P.~G.}\ \bibnamefont {Kwiat}},\
  }\href@noop {} {\bibfield  {journal} {\bibinfo  {journal} {Nature physics}\
  }\textbf {\bibinfo {volume} {4}},\ \bibinfo {pages} {282} (\bibinfo {year}
  {2008})}\BibitemShut {NoStop}%
\bibitem [{\citenamefont {Wang}\ \emph {et~al.}(2012)\citenamefont {Wang},
  \citenamefont {Yang}, \citenamefont {Fazal}, \citenamefont {Ahmed},
  \citenamefont {Yan}, \citenamefont {Huang}, \citenamefont {Ren},
  \citenamefont {Yue}, \citenamefont {Dolinar}, \citenamefont {Tur} \emph
  {et~al.}}]{wang2012terabit}%
  \BibitemOpen
  \bibfield  {author} {\bibinfo {author} {\bibfnamefont {J.}~\bibnamefont
  {Wang}}, \bibinfo {author} {\bibfnamefont {J.-Y.}\ \bibnamefont {Yang}},
  \bibinfo {author} {\bibfnamefont {I.~M.}\ \bibnamefont {Fazal}}, \bibinfo
  {author} {\bibfnamefont {N.}~\bibnamefont {Ahmed}}, \bibinfo {author}
  {\bibfnamefont {Y.}~\bibnamefont {Yan}}, \bibinfo {author} {\bibfnamefont
  {H.}~\bibnamefont {Huang}}, \bibinfo {author} {\bibfnamefont
  {Y.}~\bibnamefont {Ren}}, \bibinfo {author} {\bibfnamefont {Y.}~\bibnamefont
  {Yue}}, \bibinfo {author} {\bibfnamefont {S.}~\bibnamefont {Dolinar}},
  \bibinfo {author} {\bibfnamefont {M.}~\bibnamefont {Tur}},  \emph {et~al.},\
  }\href@noop {} {\bibfield  {journal} {\bibinfo  {journal} {Nature photonics}\
  }\textbf {\bibinfo {volume} {6}},\ \bibinfo {pages} {488} (\bibinfo {year}
  {2012})}\BibitemShut {NoStop}%
\bibitem [{\citenamefont {Ren}\ \emph {et~al.}(2016)\citenamefont {Ren},
  \citenamefont {Li}, \citenamefont {Zhang},\ and\ \citenamefont
  {Gu}}]{ren2016chip}%
  \BibitemOpen
  \bibfield  {author} {\bibinfo {author} {\bibfnamefont {H.}~\bibnamefont
  {Ren}}, \bibinfo {author} {\bibfnamefont {X.}~\bibnamefont {Li}}, \bibinfo
  {author} {\bibfnamefont {Q.}~\bibnamefont {Zhang}}, \ and\ \bibinfo {author}
  {\bibfnamefont {M.}~\bibnamefont {Gu}},\ }\href@noop {} {\bibfield  {journal}
  {\bibinfo  {journal} {Science}\ }\textbf {\bibinfo {volume} {352}},\ \bibinfo
  {pages} {805} (\bibinfo {year} {2016})}\BibitemShut {NoStop}%
\bibitem [{\citenamefont {Cai}\ \emph {et~al.}(2012)\citenamefont {Cai},
  \citenamefont {Wang}, \citenamefont {Strain}, \citenamefont {Johnson-Morris},
  \citenamefont {Zhu}, \citenamefont {Sorel}, \citenamefont {O’Brien},
  \citenamefont {Thompson},\ and\ \citenamefont {Yu}}]{cai2012integrated}%
  \BibitemOpen
  \bibfield  {author} {\bibinfo {author} {\bibfnamefont {X.}~\bibnamefont
  {Cai}}, \bibinfo {author} {\bibfnamefont {J.}~\bibnamefont {Wang}}, \bibinfo
  {author} {\bibfnamefont {M.~J.}\ \bibnamefont {Strain}}, \bibinfo {author}
  {\bibfnamefont {B.}~\bibnamefont {Johnson-Morris}}, \bibinfo {author}
  {\bibfnamefont {J.}~\bibnamefont {Zhu}}, \bibinfo {author} {\bibfnamefont
  {M.}~\bibnamefont {Sorel}}, \bibinfo {author} {\bibfnamefont {J.~L.}\
  \bibnamefont {O’Brien}}, \bibinfo {author} {\bibfnamefont {M.~G.}\
  \bibnamefont {Thompson}}, \ and\ \bibinfo {author} {\bibfnamefont
  {S.}~\bibnamefont {Yu}},\ }\href@noop {} {\bibfield  {journal} {\bibinfo
  {journal} {Science}\ }\textbf {\bibinfo {volume} {338}},\ \bibinfo {pages}
  {363} (\bibinfo {year} {2012})}\BibitemShut {NoStop}%
\bibitem [{\citenamefont {Miao}\ \emph {et~al.}(2016)\citenamefont {Miao},
  \citenamefont {Zhang}, \citenamefont {Sun}, \citenamefont {Walasik},
  \citenamefont {Longhi}, \citenamefont {Litchinitser},\ and\ \citenamefont
  {Feng}}]{miao2016orbital}%
  \BibitemOpen
  \bibfield  {author} {\bibinfo {author} {\bibfnamefont {P.}~\bibnamefont
  {Miao}}, \bibinfo {author} {\bibfnamefont {Z.}~\bibnamefont {Zhang}},
  \bibinfo {author} {\bibfnamefont {J.}~\bibnamefont {Sun}}, \bibinfo {author}
  {\bibfnamefont {W.}~\bibnamefont {Walasik}}, \bibinfo {author} {\bibfnamefont
  {S.}~\bibnamefont {Longhi}}, \bibinfo {author} {\bibfnamefont {N.~M.}\
  \bibnamefont {Litchinitser}}, \ and\ \bibinfo {author} {\bibfnamefont
  {L.}~\bibnamefont {Feng}},\ }\href@noop {} {\bibfield  {journal} {\bibinfo
  {journal} {Science}\ }\textbf {\bibinfo {volume} {353}},\ \bibinfo {pages}
  {464} (\bibinfo {year} {2016})}\BibitemShut {NoStop}%
\bibitem [{\citenamefont {Liu}\ \emph {et~al.}(2017)\citenamefont {Liu},
  \citenamefont {Zhao}, \citenamefont {Ding}, \citenamefont {Yao},
  \citenamefont {Fan},\ and\ \citenamefont {Shen}}]{liu2017wavelength}%
  \BibitemOpen
  \bibfield  {author} {\bibinfo {author} {\bibfnamefont {Q.}~\bibnamefont
  {Liu}}, \bibinfo {author} {\bibfnamefont {Y.}~\bibnamefont {Zhao}}, \bibinfo
  {author} {\bibfnamefont {M.}~\bibnamefont {Ding}}, \bibinfo {author}
  {\bibfnamefont {W.}~\bibnamefont {Yao}}, \bibinfo {author} {\bibfnamefont
  {X.}~\bibnamefont {Fan}}, \ and\ \bibinfo {author} {\bibfnamefont
  {D.}~\bibnamefont {Shen}},\ }\href@noop {} {\bibfield  {journal} {\bibinfo
  {journal} {Optics Express}\ }\textbf {\bibinfo {volume} {25}},\ \bibinfo
  {pages} {23312} (\bibinfo {year} {2017})}\BibitemShut {NoStop}%
\bibitem [{\citenamefont {Kitamura}\ \emph {et~al.}(2019)\citenamefont
  {Kitamura}, \citenamefont {Kitazawa},\ and\ \citenamefont
  {Noda}}]{kitamura2019generation}%
  \BibitemOpen
  \bibfield  {author} {\bibinfo {author} {\bibfnamefont {K.}~\bibnamefont
  {Kitamura}}, \bibinfo {author} {\bibfnamefont {M.}~\bibnamefont {Kitazawa}},
  \ and\ \bibinfo {author} {\bibfnamefont {S.}~\bibnamefont {Noda}},\
  }\href@noop {} {\bibfield  {journal} {\bibinfo  {journal} {Optics Express}\
  }\textbf {\bibinfo {volume} {27}},\ \bibinfo {pages} {1045} (\bibinfo {year}
  {2019})}\BibitemShut {NoStop}%
\bibitem [{\citenamefont {Bahari}\ \emph {et~al.}(2021)\citenamefont {Bahari},
  \citenamefont {Hsu}, \citenamefont {Pan}, \citenamefont {Preece},
  \citenamefont {Ndao}, \citenamefont {El~Amili}, \citenamefont {Fainman},\
  and\ \citenamefont {Kant{\'e}}}]{bahari2021photonic}%
  \BibitemOpen
  \bibfield  {author} {\bibinfo {author} {\bibfnamefont {B.}~\bibnamefont
  {Bahari}}, \bibinfo {author} {\bibfnamefont {L.}~\bibnamefont {Hsu}},
  \bibinfo {author} {\bibfnamefont {S.~H.}\ \bibnamefont {Pan}}, \bibinfo
  {author} {\bibfnamefont {D.}~\bibnamefont {Preece}}, \bibinfo {author}
  {\bibfnamefont {A.}~\bibnamefont {Ndao}}, \bibinfo {author} {\bibfnamefont
  {A.}~\bibnamefont {El~Amili}}, \bibinfo {author} {\bibfnamefont
  {Y.}~\bibnamefont {Fainman}}, \ and\ \bibinfo {author} {\bibfnamefont
  {B.}~\bibnamefont {Kant{\'e}}},\ }\href@noop {} {\bibfield  {journal}
  {\bibinfo  {journal} {Nature Physics}\ }\textbf {\bibinfo {volume} {17}},\
  \bibinfo {pages} {700} (\bibinfo {year} {2021})}\BibitemShut {NoStop}%
\bibitem [{\citenamefont {Yu}\ \emph {et~al.}(2011)\citenamefont {Yu},
  \citenamefont {Genevet}, \citenamefont {Kats}, \citenamefont {Aieta},
  \citenamefont {Tetienne}, \citenamefont {Capasso},\ and\ \citenamefont
  {Gaburro}}]{yu2011light}%
  \BibitemOpen
  \bibfield  {author} {\bibinfo {author} {\bibfnamefont {N.}~\bibnamefont
  {Yu}}, \bibinfo {author} {\bibfnamefont {P.}~\bibnamefont {Genevet}},
  \bibinfo {author} {\bibfnamefont {M.~A.}\ \bibnamefont {Kats}}, \bibinfo
  {author} {\bibfnamefont {F.}~\bibnamefont {Aieta}}, \bibinfo {author}
  {\bibfnamefont {J.-P.}\ \bibnamefont {Tetienne}}, \bibinfo {author}
  {\bibfnamefont {F.}~\bibnamefont {Capasso}}, \ and\ \bibinfo {author}
  {\bibfnamefont {Z.}~\bibnamefont {Gaburro}},\ }\href@noop {} {\bibfield
  {journal} {\bibinfo  {journal} {science}\ }\textbf {\bibinfo {volume}
  {334}},\ \bibinfo {pages} {333} (\bibinfo {year} {2011})}\BibitemShut
  {NoStop}%
\bibitem [{\citenamefont {Chen}\ \emph {et~al.}(2018)\citenamefont {Chen},
  \citenamefont {Jiang},\ and\ \citenamefont {Sha}}]{chen2018orbital}%
  \BibitemOpen
  \bibfield  {author} {\bibinfo {author} {\bibfnamefont {M.~L.}\ \bibnamefont
  {Chen}}, \bibinfo {author} {\bibfnamefont {L.~J.}\ \bibnamefont {Jiang}}, \
  and\ \bibinfo {author} {\bibfnamefont {W.~E.}\ \bibnamefont {Sha}},\
  }\href@noop {} {\bibfield  {journal} {\bibinfo  {journal} {Applied Sciences}\
  }\textbf {\bibinfo {volume} {8}},\ \bibinfo {pages} {362} (\bibinfo {year}
  {2018})}\BibitemShut {NoStop}%
\bibitem [{\citenamefont {Sedeh}\ \emph {et~al.}(2020)\citenamefont {Sedeh},
  \citenamefont {Salary},\ and\ \citenamefont {Mosallaei}}]{sedeh2020time}%
  \BibitemOpen
  \bibfield  {author} {\bibinfo {author} {\bibfnamefont {H.~B.}\ \bibnamefont
  {Sedeh}}, \bibinfo {author} {\bibfnamefont {M.~M.}\ \bibnamefont {Salary}}, \
  and\ \bibinfo {author} {\bibfnamefont {H.}~\bibnamefont {Mosallaei}},\
  }\href@noop {} {\bibfield  {journal} {\bibinfo  {journal} {Nanophotonics}\
  }\textbf {\bibinfo {volume} {9}},\ \bibinfo {pages} {2957} (\bibinfo {year}
  {2020})}\BibitemShut {NoStop}%
\bibitem [{\citenamefont {Su}\ \emph {et~al.}(2012)\citenamefont {Su},
  \citenamefont {Scott}, \citenamefont {Djordjevic}, \citenamefont {Fontaine},
  \citenamefont {Geisler}, \citenamefont {Cai},\ and\ \citenamefont
  {Yoo}}]{su2012demonstration}%
  \BibitemOpen
  \bibfield  {author} {\bibinfo {author} {\bibfnamefont {T.}~\bibnamefont
  {Su}}, \bibinfo {author} {\bibfnamefont {R.~P.}\ \bibnamefont {Scott}},
  \bibinfo {author} {\bibfnamefont {S.~S.}\ \bibnamefont {Djordjevic}},
  \bibinfo {author} {\bibfnamefont {N.~K.}\ \bibnamefont {Fontaine}}, \bibinfo
  {author} {\bibfnamefont {D.~J.}\ \bibnamefont {Geisler}}, \bibinfo {author}
  {\bibfnamefont {X.}~\bibnamefont {Cai}}, \ and\ \bibinfo {author}
  {\bibfnamefont {S.}~\bibnamefont {Yoo}},\ }\href@noop {} {\bibfield
  {journal} {\bibinfo  {journal} {Optics express}\ }\textbf {\bibinfo {volume}
  {20}},\ \bibinfo {pages} {9396} (\bibinfo {year} {2012})}\BibitemShut
  {NoStop}%
\bibitem [{\citenamefont {Zhou}\ \emph {et~al.}(2019)\citenamefont {Zhou},
  \citenamefont {Zheng}, \citenamefont {Cao}, \citenamefont {Zhao},
  \citenamefont {Gao}, \citenamefont {Zhu}, \citenamefont {He}, \citenamefont
  {Cai},\ and\ \citenamefont {Wang}}]{zhou2019ultra}%
  \BibitemOpen
  \bibfield  {author} {\bibinfo {author} {\bibfnamefont {N.}~\bibnamefont
  {Zhou}}, \bibinfo {author} {\bibfnamefont {S.}~\bibnamefont {Zheng}},
  \bibinfo {author} {\bibfnamefont {X.}~\bibnamefont {Cao}}, \bibinfo {author}
  {\bibfnamefont {Y.}~\bibnamefont {Zhao}}, \bibinfo {author} {\bibfnamefont
  {S.}~\bibnamefont {Gao}}, \bibinfo {author} {\bibfnamefont {Y.}~\bibnamefont
  {Zhu}}, \bibinfo {author} {\bibfnamefont {M.}~\bibnamefont {He}}, \bibinfo
  {author} {\bibfnamefont {X.}~\bibnamefont {Cai}}, \ and\ \bibinfo {author}
  {\bibfnamefont {J.}~\bibnamefont {Wang}},\ }\href@noop {} {\bibfield
  {journal} {\bibinfo  {journal} {Science advances}\ }\textbf {\bibinfo
  {volume} {5}},\ \bibinfo {pages} {eaau9593} (\bibinfo {year}
  {2019})}\BibitemShut {NoStop}%
\bibitem [{\citenamefont {Zhao}\ and\ \citenamefont
  {Fan}(2020)}]{zhao2020design}%
  \BibitemOpen
  \bibfield  {author} {\bibinfo {author} {\bibfnamefont {Z.}~\bibnamefont
  {Zhao}}\ and\ \bibinfo {author} {\bibfnamefont {S.}~\bibnamefont {Fan}},\
  }\href@noop {} {\bibfield  {journal} {\bibinfo  {journal} {Journal of
  Lightwave Technology}\ }\textbf {\bibinfo {volume} {38}},\ \bibinfo {pages}
  {4435} (\bibinfo {year} {2020})}\BibitemShut {NoStop}%
\bibitem [{\citenamefont {Xie}\ \emph {et~al.}(2018)\citenamefont {Xie},
  \citenamefont {Lei}, \citenamefont {Li}, \citenamefont {Qiu}, \citenamefont
  {Zhang}, \citenamefont {Wang}, \citenamefont {Min}, \citenamefont {Du},
  \citenamefont {Li},\ and\ \citenamefont {Yuan}}]{xie2018ultra}%
  \BibitemOpen
  \bibfield  {author} {\bibinfo {author} {\bibfnamefont {Z.}~\bibnamefont
  {Xie}}, \bibinfo {author} {\bibfnamefont {T.}~\bibnamefont {Lei}}, \bibinfo
  {author} {\bibfnamefont {F.}~\bibnamefont {Li}}, \bibinfo {author}
  {\bibfnamefont {H.}~\bibnamefont {Qiu}}, \bibinfo {author} {\bibfnamefont
  {Z.}~\bibnamefont {Zhang}}, \bibinfo {author} {\bibfnamefont
  {H.}~\bibnamefont {Wang}}, \bibinfo {author} {\bibfnamefont {C.}~\bibnamefont
  {Min}}, \bibinfo {author} {\bibfnamefont {L.}~\bibnamefont {Du}}, \bibinfo
  {author} {\bibfnamefont {Z.}~\bibnamefont {Li}}, \ and\ \bibinfo {author}
  {\bibfnamefont {X.}~\bibnamefont {Yuan}},\ }\href@noop {} {\bibfield
  {journal} {\bibinfo  {journal} {Light: Science \& Applications}\ }\textbf
  {\bibinfo {volume} {7}},\ \bibinfo {pages} {18001} (\bibinfo {year}
  {2018})}\BibitemShut {NoStop}%
\bibitem [{\citenamefont {Song}\ \emph {et~al.}(2020)\citenamefont {Song},
  \citenamefont {Zhao}, \citenamefont {Zhang}, \citenamefont {Song},
  \citenamefont {Zhou}, \citenamefont {Pang}, \citenamefont {Du}, \citenamefont
  {Li}, \citenamefont {Liu}, \citenamefont {Su} \emph
  {et~al.}}]{song2020utilizing}%
  \BibitemOpen
  \bibfield  {author} {\bibinfo {author} {\bibfnamefont {H.}~\bibnamefont
  {Song}}, \bibinfo {author} {\bibfnamefont {Z.}~\bibnamefont {Zhao}}, \bibinfo
  {author} {\bibfnamefont {R.}~\bibnamefont {Zhang}}, \bibinfo {author}
  {\bibfnamefont {H.}~\bibnamefont {Song}}, \bibinfo {author} {\bibfnamefont
  {H.}~\bibnamefont {Zhou}}, \bibinfo {author} {\bibfnamefont {K.}~\bibnamefont
  {Pang}}, \bibinfo {author} {\bibfnamefont {J.}~\bibnamefont {Du}}, \bibinfo
  {author} {\bibfnamefont {L.}~\bibnamefont {Li}}, \bibinfo {author}
  {\bibfnamefont {C.}~\bibnamefont {Liu}}, \bibinfo {author} {\bibfnamefont
  {X.}~\bibnamefont {Su}},  \emph {et~al.},\ }\href@noop {} {\bibfield
  {journal} {\bibinfo  {journal} {Optics Letters}\ }\textbf {\bibinfo {volume}
  {45}},\ \bibinfo {pages} {4144} (\bibinfo {year} {2020})}\BibitemShut
  {NoStop}%
\bibitem [{\citenamefont {White}\ \emph {et~al.}(2021)\citenamefont {White},
  \citenamefont {Yang},\ and\ \citenamefont
  {Vu{\v{c}}kovi{\'c}}}]{white2021inverse}%
  \BibitemOpen
  \bibfield  {author} {\bibinfo {author} {\bibfnamefont {A.}~\bibnamefont
  {White}}, \bibinfo {author} {\bibfnamefont {K.}~\bibnamefont {Yang}}, \ and\
  \bibinfo {author} {\bibfnamefont {J.}~\bibnamefont {Vu{\v{c}}kovi{\'c}}},\
  }in\ \href@noop {} {\emph {\bibinfo {booktitle} {CLEO: Science and
  Innovations}}}\ (\bibinfo {organization} {Optical Society of America},\
  \bibinfo {year} {2021})\ pp.\ \bibinfo {pages} {SM4C--2}\BibitemShut
  {NoStop}%
\bibitem [{\citenamefont {Bauer}\ \emph {et~al.}(2015)\citenamefont {Bauer},
  \citenamefont {Banzer}, \citenamefont {Karimi}, \citenamefont {Orlov},
  \citenamefont {Rubano}, \citenamefont {Marrucci}, \citenamefont {Santamato},
  \citenamefont {Boyd},\ and\ \citenamefont {Leuchs}}]{bauer2015observation}%
  \BibitemOpen
  \bibfield  {author} {\bibinfo {author} {\bibfnamefont {T.}~\bibnamefont
  {Bauer}}, \bibinfo {author} {\bibfnamefont {P.}~\bibnamefont {Banzer}},
  \bibinfo {author} {\bibfnamefont {E.}~\bibnamefont {Karimi}}, \bibinfo
  {author} {\bibfnamefont {S.}~\bibnamefont {Orlov}}, \bibinfo {author}
  {\bibfnamefont {A.}~\bibnamefont {Rubano}}, \bibinfo {author} {\bibfnamefont
  {L.}~\bibnamefont {Marrucci}}, \bibinfo {author} {\bibfnamefont
  {E.}~\bibnamefont {Santamato}}, \bibinfo {author} {\bibfnamefont {R.~W.}\
  \bibnamefont {Boyd}}, \ and\ \bibinfo {author} {\bibfnamefont
  {G.}~\bibnamefont {Leuchs}},\ }\href@noop {} {\bibfield  {journal} {\bibinfo
  {journal} {Science}\ }\textbf {\bibinfo {volume} {347}},\ \bibinfo {pages}
  {964} (\bibinfo {year} {2015})}\BibitemShut {NoStop}%
\bibitem [{\citenamefont {Goos}(2021)}]{goos}%
  \BibitemOpen
  \bibfield  {author} {\bibinfo {author} {\bibnamefont {Goos}},\ }\href@noop {}
  {}\bibinfo {howpublished} {\url{https://github.com/stanfordnqp/spins-b}}
  (\bibinfo {year} {2021})\BibitemShut {NoStop}%
\bibitem [{\citenamefont {Su}\ \emph {et~al.}(2020)\citenamefont {Su},
  \citenamefont {Vercruysse}, \citenamefont {Skarda}, \citenamefont {Sapra},
  \citenamefont {Petykiewicz},\ and\ \citenamefont
  {Vu{\v{c}}kovi{\'c}}}]{su2020nanophotonic}%
  \BibitemOpen
  \bibfield  {author} {\bibinfo {author} {\bibfnamefont {L.}~\bibnamefont
  {Su}}, \bibinfo {author} {\bibfnamefont {D.}~\bibnamefont {Vercruysse}},
  \bibinfo {author} {\bibfnamefont {J.}~\bibnamefont {Skarda}}, \bibinfo
  {author} {\bibfnamefont {N.~V.}\ \bibnamefont {Sapra}}, \bibinfo {author}
  {\bibfnamefont {J.~A.}\ \bibnamefont {Petykiewicz}}, \ and\ \bibinfo {author}
  {\bibfnamefont {J.}~\bibnamefont {Vu{\v{c}}kovi{\'c}}},\ }\href@noop {}
  {\bibfield  {journal} {\bibinfo  {journal} {Applied Physics Reviews}\
  }\textbf {\bibinfo {volume} {7}},\ \bibinfo {pages} {011407} (\bibinfo {year}
  {2020})}\BibitemShut {NoStop}%
\bibitem [{\citenamefont {Bozinovic}\ \emph {et~al.}(2013)\citenamefont
  {Bozinovic}, \citenamefont {Yue}, \citenamefont {Ren}, \citenamefont {Tur},
  \citenamefont {Kristensen}, \citenamefont {Huang}, \citenamefont {Willner},\
  and\ \citenamefont {Ramachandran}}]{bozinovic2013terabit}%
  \BibitemOpen
  \bibfield  {author} {\bibinfo {author} {\bibfnamefont {N.}~\bibnamefont
  {Bozinovic}}, \bibinfo {author} {\bibfnamefont {Y.}~\bibnamefont {Yue}},
  \bibinfo {author} {\bibfnamefont {Y.}~\bibnamefont {Ren}}, \bibinfo {author}
  {\bibfnamefont {M.}~\bibnamefont {Tur}}, \bibinfo {author} {\bibfnamefont
  {P.}~\bibnamefont {Kristensen}}, \bibinfo {author} {\bibfnamefont
  {H.}~\bibnamefont {Huang}}, \bibinfo {author} {\bibfnamefont {A.~E.}\
  \bibnamefont {Willner}}, \ and\ \bibinfo {author} {\bibfnamefont
  {S.}~\bibnamefont {Ramachandran}},\ }\href@noop {} {\bibfield  {journal}
  {\bibinfo  {journal} {Science}\ }\textbf {\bibinfo {volume} {340}},\ \bibinfo
  {pages} {1545} (\bibinfo {year} {2013})}\BibitemShut {NoStop}%
\bibitem [{\citenamefont {Sacher}\ \emph {et~al.}(2015)\citenamefont {Sacher},
  \citenamefont {Huang}, \citenamefont {Lo},\ and\ \citenamefont
  {Poon}}]{sacher2015multilayer}%
  \BibitemOpen
  \bibfield  {author} {\bibinfo {author} {\bibfnamefont {W.~D.}\ \bibnamefont
  {Sacher}}, \bibinfo {author} {\bibfnamefont {Y.}~\bibnamefont {Huang}},
  \bibinfo {author} {\bibfnamefont {G.-Q.}\ \bibnamefont {Lo}}, \ and\ \bibinfo
  {author} {\bibfnamefont {J.~K.}\ \bibnamefont {Poon}},\ }\href@noop {}
  {\bibfield  {journal} {\bibinfo  {journal} {Journal of lightwave technology}\
  }\textbf {\bibinfo {volume} {33}},\ \bibinfo {pages} {901} (\bibinfo {year}
  {2015})}\BibitemShut {NoStop}%
\end{thebibliography}%
